\documentstyle[12pt]{article}
\author{Hans - J\"urgen Schmidt}
\title{Consequences of the noncompactness of the
Lorentz group}
\date{}
\begin{document}
\maketitle

\bigskip

\centerline{
Universit\"at  Potsdam, Institut f\"ur Mathematik}
\centerline{  Projektgruppe
Kosmologie}
\centerline{
      D-14415 POTSDAM, PF 601553, Am Neuen Palais 10, Germany}

\bigskip

\begin{abstract}

The following four statements have been proven decades ago
already, but they continue to induce a strange feeling:

- All curvature invariants of a gravitational wave
vanish - inspite of the fact that it represents a
nonflat spacetime.

- The eigennullframe components of the curvature
tensor (the Cartan ''scalars'') do not represent
curvature scalars.

- The Euclidean topology in the Minkowski spacetime
does not possess a basis composed of Lorentz--invariant
neighbourhoods.

- There are points in the de Sitter spacetime which
 cannot be joined to each other by any geodesic.

We explain that our feeling is influenced by the
compactness of the rotation group; the strangeness
disappears if we fully acknowledge the noncompactness
of the Lorentz group.

\noindent
Output: Imaginary coordinate rotations from Euclidean
 to Lorentzian signature are very dangerous.

\end{abstract}

AMS number: 53 B 30 Lorentz metrics, indefinite metrics


\section{ Introduction }

A topological space $X$ is compact iff each open
cover contains a finite subcover. Equivalently one
can say: $X$ is compact iff each sequence in $X$
possesses a converging subsequence.
$SO(n)$, the $n$--dimensional rotation group, is compact,
whereas $SO(n-1,1)$, the corresponding Lorentz group,
fails to be compact. Nevertheless, one can simply
switch from the Euclidean space $E^n$ to the
Minkowski spacetime $M^n$ by replacing
$x^n \longrightarrow it$.

It is the aim of the present essay to show those points
where the loss of compactness connected with this
replacement has nontrivial consequences.

\section{Gravitational waves}
\setcounter{equation}{0}

Let
\begin{equation}
ds \sp 2 \quad = \quad 2 \, du \, dv \, - \, a\sp 2 (u)
 \, dw \sp 2 \, - \, b\sp 2 (u) \, dz \sp 2
\end{equation}
with positive smooth functions $a$ and $b$.
It represents a gravitational wave iff
\begin{equation}
 a \cdot \frac{d^2b}{du^2} \quad + \quad
  b \cdot \frac{d^2a}{du^2} \quad = \quad 0
\end{equation}
cf. e.g. the review [1]. Metric (2.1) represents a flat
spacetime iff both $a$ and $b$ are linear functions.

Let $I$ be any curvature  invariant of order $k$,
i.e.,
{\large
$$
 I = I(g_{ij},  \,  R_{ijlm}, \dots ,
 R_{ijlm;i_1 \dots \, i_k} )
$$
}
is a scalar
 which depends continuously on  all  its   arguments;
 the domain of dependence is requested to contain  the
flat space, and $ I(g_{ij}, \, 0, \dots \, 0) \equiv 0 $.

It holds, cf. [2, 3]: For gravitational waves of type (2.1),
$I$ identically vanishes. Moreover, one can prove that
statement for all metrics (2.1) without requiring (2.2).
The proof by calculating the components of the curvature
tensor is possible but quite technical.

	A very short and geometrical proof goes as follows:
Apply a Lorentz boost in the $u-v-$plane,
i.e., $u \longrightarrow \lambda \cdot u$ and
$v \longrightarrow \lambda ^{-1} \cdot v$ for any
$\lambda > 0$. Then $a(u)$ is replaced by
$a(\lambda \cdot u)$ and $b(u)$ by
$b(\lambda \cdot u)$. In the limit
$\lambda \longrightarrow 0$,
metric (2.1) has a unique limit: constant functions $a$
and $b$. It is the  flat spacetime, and so, $I=0$ there.
On the other hand, for all positive values $\lambda$,
$I$ carries the same value. By continuity this value
equals zero. q.e.d.

Why all trials failed to generalize the idea of this proof
to the positive definite case? Because we need
a sequence of Lorentz boosts which does not
possess any accumulation points within $SO(3,1)$.
Such a sequence does not exist in $SO(4)$
because of compactness.

\section{Cartan scalars}
\setcounter{equation}{0}

A variant [4] of the Newman--Penrose formalism
uses projections to an eigennullframe of the
curvature tensor to classify gravitational waves.
The corresponding Cartan ''scalars'' have
different boost weights, and they represent curvature
 invariants for vanishing boost weight only.
 The non--vanishing Cartan ''scalars'' for metric (2.1)
have either non--vanishing boost weight or a
discontinuity at flat spacetime. So, the GHP-formalism
 [4], cf. also [5],
does not yield a contradiction to the statement in sct. 2.

Why do not exist analogous ''scalars'' with different
''rotation weight'' in the Euclidean signature case?
Analysing the construction one can see:
The boost weights appear because there is a nontrivial vector
space isomorphic to a closed subgroup of $SO(3,1)$.
For $SO(4)$, however, it holds:
Every closed subgroup is compact; in order that it is
isomorphic to a vector space it is necessary that it is
the trivial one--point space.

More geometrically this looks as follows, cf. [6]:
 Let $v \in E^4$ be a vector and $g \in SO(4)$ such that
$g(v) \uparrow \uparrow v$, then it holds: $g(v)=v$.
In Minkowski
space--time $M^4$, however, there exist vectors
 $v \in M^4$
 and boosts $h \in SO(3,1)$ with
 $h(v)  \uparrow \uparrow v$ and
$h(v) \ne v$.

\section{Lorentz--invariant neighbourhoods}
\setcounter{equation}{0}

There is no doubt that the Euclidean topology $\tau$
is the adequate topology of the Euclidean space $E^n$.
 However, controversies appear if one asks
whether
 $\tau$ is best suited for the Minkowski spacetime
$M^n$.

	The most radical path in answering this
question can be found in refs. [7, 8, 9]; it leads
to a topology different from  $\tau$ which
fails to be a normal one.

	Here, we only want to find out in which sense
one can say that  $\tau$ is better adapted to $E^n$ than
to $M^n$. From a first view they appear on an
equal footing: Both $SO(n)$ and $SO(n-1,1)$
represent subgroups of the homeomorphism group of  $\tau$.

	The difference appears as follows:
For $E^n$, the
usual $\epsilon-$spheres form a neighbourhood--basis
composed of $SO(n)$-invariant open sets. Moreover,
each of these neighbourhoods has a compact closure.
 Let $U$ be any open neighbourhood with
compact closure around  the origin in $E^n$.
For every $g \in SO(n)$, $U(g)$ is the set $U$ after
rotation by $g$. Of course,  $U(g)$ is also an
open neighbourhood with
compact closure around  the origin in $E^n$.
Let us define
$$
V \quad = \quad \bigcup \{ U(g) \vert g \in SO(n)\}
$$
and
$$
W \quad = \quad \bigcap \{ U(g) \vert g \in SO(n)\}
$$
It holds: Both $V$ and $W$ represent
  $SO(n)$-invariant
neighbourhoods of the origin with compact closure.
Analysing the proofs one can see: ''$V$
is a neighbourhood of the origin''
and ''$W$ has compact closure'' are trivial
statements, whereas
''$W$
is a neighbourhood of the origin''
and ''$V$ has compact closure''
essentially need the compactness of $SO(n)$.
For $M^n$, however, all these properties fail.

First: No point of $M^n$ possesses a  neighbourhood--basis
composed of $SO(n-1,1)$-invariant open sets.

Second:  No $SO(n-1,1)$-invariant neighbourhood
has a compact closure.

Third:  Let $U$ be any open neighbourhood with
compact closure around  the origin in $M^n$.
For every $g \in SO(n-1,1)$,  $U(g)$ is also an
open neighbourhood with
compact closure around  the origin in $M^n$.
However, neither
$$
V \quad = \quad \bigcup \{ U(g) \vert g \in SO(n-1,1)\}
$$
nor
$$
W \quad = \quad \bigcap \{ U(g) \vert g \in SO(n-1,1) \}
$$
represent
neighbourhoods of the origin with compact closure.

\section{Geodesics}
\setcounter{equation}{0}

Now we analyze a statement (known already to
 de Sitter himself, cf. [2, 10]):
Inspite of the fact that the de Sitter spacetime is
connected and geodetically complete,
there are points in it  which
 cannot be joined to each other by any geodesic.

Let us recall: For Riemannian spaces it holds: If the
space is connected and geodetically complete, then
each pair of points can be connected by a geodesic.

The proof for Riemannian spaces $V_n$ goes as follows:
Take one of its points as $x$ and define
$M_x \subset V_n$
to be that  set of points which can be
reached from $x$ by a geodesic. One can show that
$M_x$ is non--empty, open and closed. This
implies $M_x = V_n$.

But where does the corresponding proof fail
 when we try to generalize it to the de Sitter
spacetime?

Let us recall: A geodetic $\epsilon$--ball is the
exponentiated form of a rotation--invariant
neighbourhood of the corresponding tangent space.
For Riemannian spaces these geodetic $\epsilon$--balls
form a neighbourhood basis - and just this is needed
in the proof.

But where does it fail in detail ?  $M_x$  ''non--empty,
open and closed'' would again imply $M_x = V_n$.
$M_x$  ''non--empty'' is trivially satisfied by
$x \in  M_x$. So we can fail by proving ''open''
or by proving ''closed''. It turns out, cf. [10],
that $M_x$ is neither open nor closed, and
both properties fail by the lack of a neighbourhood
basis consisting of geodetic $\epsilon$--balls.

So, if compared with sct. 4, we can see that it is again
the noncompactness of the Lorentz group which
 produces  the peculiarities.

\section{Conclusion}
\setcounter{equation}{0}

A finite set in set theory, a bounded set
in geometry, and a compact set in topology: these
are corresponding fundamental notions.

What have we learned from the above analysis
 on compactness?
Let us concentrate on the first point (sct. 2):
The fact that non--isometric spacetimes exist
which cannot be distinguished by curvature invariants
is neither connected with the fact that one
of them is flat nor with the vanishing of
the curvature invariants, but, as we have seen, with
the appearance of a Lorentz boost which
has a limit not belonging to $SO(3,1)$
but producing a regular
metric there. So we have found  the very recipe
to construct several classes of such spacetimes.
Let us present one of them [6]:

For a positive $C^{\infty}$--function $a(u)$ let
{\large
$$ ds^2 \ = \ \frac{1}{z^2} [ 2 \, du \, dv
\ - \  a^2(u) \, dy^2 \ - \ dz^2 ]$$}
In the region $z>0$, $ds^2$ represents the anti-de Sitter
space--time if and only
if $a(u)$ is linear in $u$. Now, let
$d^2a/du^2 \, < \, 0$ and
$$ \phi \ := \ \frac{1}{\sqrt{\kappa}} \int
\left( - \frac{1}{a} \, \frac{d^2a}{du^2} \right) ^{1/2}
 \, du $$
Then $\Box \phi \, = \,  \phi_{,i} \, \phi^{,i} \, = \, 0$
and $R_{ij} \ - \ \frac{R}{2} \, g_{ij} \ = \ \Lambda \,
g_{ij} \ + \ \kappa \, T_{ij}$ with
$\Lambda \, = \, - 3$ and
$ T_{ij} \, = \, \phi_{,i} \, \phi_{,j}$.
So $(ds^2, \, \phi)$ represents a solution of Einstein's
equation with negative cosmological term $\Lambda $ and
a minimally coupled massless scalar field $\phi $.
Let $I$ be a curvature invariant of order $k$. Then for
 the metric $ds^2$, $I$ does not depend on the
 function $a(u)$. So $I$ takes the same value both
for linear and non--linear functions $a(u)$.
This seems to be the first example that
 non--isometric
space--times with non--vanishing curvature scalar
 cannot be distinguished by curvature invariants. And
having the recipe the construction of other classes
is straightforwardly done.

The fact that the representation theory of the
rotation groups
$SO(n)$ and the Lorentz groups $SO(n-1,1)$
 is quite different is so well--known that we
did not repeat it here - we only want to mention that
 it is the  compactness
of the first one which produces the
difference.

\bigskip

{\Large
{\bf Acknowledgement}}

\bigskip

Financial support from the DFG
and valuable comments by Alan Held
and Martin Rainer are gratefully acknowledged.

\newpage

{\Large
{\bf References}}

\noindent
[1] Schimming, R. (1974). {\it Math. Nachr.} {\bf 59}, 129.

\noindent
[2] Hawking, S., Ellis, G. F. R. (1973) {\it The large
scale structure of space--time} (Cambridge Univ. Press).

\noindent
[3] Jordan, P.,  Ehlers, J.,  Kundt, W. (1960) {\it Abh.
Akad. Wiss. Mainz, Math./Nat.} 21.

\noindent
[4] Geroch, R., Held, A., Penrose, R. (1973) {\it J.
Math. Phys.} {\bf 14}, 874.

\noindent
[5] Dautcourt, G. et al. (1981) {\it Astron. Nachr.}
{\bf 302}, 1.

\noindent
[6] Schmidt, H.-J. (1995) {\it Abstracts GR 14
 Florence} and
{\it New frontiers in gravitation} eds.:
G. Sardanashvili, R. Santilli.

\noindent
[7] Hawking, S., King, A., McCarthy, P. (1976)
{\it J. Math. Phys.} {\bf 17}, 174.

\noindent
[8] Schmidt, H.-J. (1984) {\it Abh. Akademie
d. Wiss. d. DDR, Abt. Math.} {\bf 2N}, 207.

\noindent
[9] Fullwood, D. (1992)
{\it J. Math. Phys.} {\bf 33}, 2232.

\noindent
[10] Schmidt, H.-J. (1993) {\it Fortschr. Phys.}
 {\bf 41}, 179.

\end{document}